\RequirePackage{fix-cm}
\documentclass[twocolumn,epjc3]{svjour3}  
\smartqed  

\RequirePackage{graphicx}
\RequirePackage[caption=false]{subfig}
\RequirePackage[dvipsnames]{xcolor}
\RequirePackage{soul}
\usepackage{units} 
\newcommand{\beq}{\begin{equation}}
\newcommand{\eeq}{\end{equation}}
\newcommand{\akt}{anti-$k_T$ }
\newcommand{\kt}{$k_T$ }
\newcommand{\pt}{$p_T$ }
\newcommand{\ca}{C/A }
\newcommand{\ta}{$\tau$ }
\newcommand{\taf}{$\tau_{form}$ }
\newcommand{\eqref}[1]{(\ref{#1})}
\newcommand{\jwl}{JEWEL }
\newcommand{\pythia}{PYTHIA\,8 }
\newcommand{\pp}{$pp$ }
\newcommand{\zcut}{$z_{cut}$ }

\journalname{Eur. Phys. J. C}

\begin{document}

\title{\hfill \textnormal{\large LU TP 20-52} \\
Time reclustering for jet quenching studies 
}


\author{Liliana Apolin\'{a}rio \thanksref{e1,addr1,addr2}
        \and
        Andr\'{e} Cordeiro\thanksref{e2,addr2} 
        \and
        Korinna Zapp\thanksref{e3,addr3}
}

\thankstext{e1}{e-mail: liliana@lip.pt}
\thankstext{e2}{e-mail: andre.cordeiro@tecnico.ulisboa.pt}
\thankstext{e3}{e-mail: korinna.zapp@thep.lu.se}


\institute{
LIP, Av. Prof. Gama Pinto, 2, P-1649-003 Lisbon, Portugal \label{addr1}
\and
Instituto Superior T\'{e}cnico (IST), Universidade de Lisboa, Av. Rovisco Pais 1, 1049-001, Lisbon, Portugal \label{addr2}
\and 
Department of Astronomy and Theoretical Physics,
Lund University,
S\"olvegatan 14A,
SE-223 62 Lund, Sweden \label{addr3}}

\date{Received: date / Accepted: date}

\maketitle

\begin{abstract}
The physics program of ultra-relativistic heavy-ion collisions at the Large Hadron Collider (LHC) and Relativistic Heavy-Ion Collider (RHIC) has brought a unique insight into the hot and dense QCD matter created in such collisions, the Quark-Gluon Plasma (QGP). Jet quenching, a collection of medium-induced modifications of the jets' internal structure that occur through their development in dense QCD matter, has a unique potential to assess the time structure of the produced medium. In this work, we perform an exploratory study to identify jet reclustering tools that can potentiate future QGP tomographic measurements with jets at current energies. Our results show that by using the inverse of formation time to obtain the jet clustering history, one can identify more accurately the time structure of QCD emissions inside jets, even in the presence of jet quenching.
\keywords{Parton shower \and Jet Quenching \and Monte Carlo Event Generators}
\end{abstract}


\section{Introduction}
\label{sec:intro}

Modifications of jets in heavy-ion collisions due to interactions with the dense medium created in such reactions lead both to a suppression of the jet cross-section and alterations of their inner structure, collectively denoted as jet quenching. Jets are complex objects that build up a characteristic structure through successive emissions while they propagate in and interact with the background medium. In the absence of such a medium, this scale evolution is calculable in perturbation theory. It is responsible for the fact that jets carry information from a broad range of scales (from the hard scale set by the hard scattering matrix elements down to the hadronic scale). As jets propagate through the evolving medium, they probe the medium also at different times and positions. The spatio-temporal structure of the background thus gets imprinted together with the scale information on the modified jets. Jets thus offer a unique chance of accessing information on the scale dependence as well as the spatial profile and time evolution of the background medium. Decoding this information from the observed jets is, however, a highly non-trivial task. The reasons for this are manifold and range from the lack of analytical control to the complications of large fluctuations in the radiation pattern. In this paper, we present a first step towards accessing time information by using a non-standard clustering algorithm and give a first example how this new technique can be used to gain information about the time evolution of the medium.

Jet algorithms are the attempt to identify the spray of final state hadrons that result from the fragmentation of a high energy parton (quark or gluon) (see \cite{Salam:2009jx} for a review). The sequential recombination jet algorithm family is massively used in both proton-proton ($pp$) and heavy-ion collisions for a multitude of studies. They start by identifying the pair of particles $(i, j)$, that are closest in a distance measure given by:
\beq
d_{ij} = \min (p_{Ti}^{2p}, p_{Tj}^{2p}) \frac{\Delta R_{ij}^2}{R^2} \, ,
\label{eq:dij}
\eeq
where
\beq
\Delta R_{ij} = \sqrt{(\phi_i - \phi_j)^2 + (y_i - y_j)^2} \, ,
\eeq
with $p_{Ti(j)}, y_{i(j)}, \phi_{i(j)}$ the transverse momentum, rapidity and azimuthal angle, respectively, of the particle $i(j)$. $R$ is the jet radius that defines the maximum $(y,\phi)$ reach of the algorithm, and $p$ a continuous parameter that specifies the jet clustering algorithm. We recover the Cambridge/Aachen (C/A) \cite{Dokshitzer:1997in,Wobisch:1998wt}, \akt\cite{Cacciari:2008gp} and \kt\cite{Catani:1993hr,Ellis:1993tq} jet algorithms by setting $p = 0$, $-1$ and $1$ respectively.

The \ca and \akt algorithms play a crucial role in jet studies. The latter, typically yielding circular jets and most sensitive to high transverse momentum particles, is usually employed in the identification of \textit{signal} jets of interest. The former is a purely geometric jet algorithm, that starts clustering particles that are close in $(y,\phi)$, and progresses to larger distances. As QCD vacuum emission follows, at leading logarithmic accuracy, angular ordering, i.e., subsequent emissions are restrained from being emitted at larger angles than the previous one, the clustering sequence obtained with the \ca algorithm resembles the QCD radiation pattern.

Since the pioneering work of~\cite{Butterworth:2008iy}, which suggested to use the information about the internal structure of jets obtained from the jet clustering sequence to identify boosted Higgs bosons decaying hadronically, the study of jet substructure in a variety of contexts has become a very active topic. The general strategy is to identify jets using the \akt algorithm and, in a second step, recluster the particles forming jets with a different algorithm, typically \ca. This procedure is often combined with grooming techniques designed to reduce contamination from uncorrelated underlying event activity mostly manifest in soft large-angle structures inside jets. One such procedure is SoftDrop~\cite{Larkoski:2014wba}, which reclusters jets with the \ca algorithm and subsequently goes backwards through the clustering sequence discarding sub-jets that are soft and/or at large angle. In pp collisions the energy sharing in the first soft-drop unclustering step was shown to be a proxy to the QCD Altarelli-Parisi splitting functions~\cite{Larkoski:2015lea}.

These techniques, developed in \pp collisions, were heavily imported to the heavy-ion field, opening a new window of exploration to deepen our understanding of jet quenching mechanisms (cf. e.g. \cite{Sirunyan:2017bsd,Caucal:2019uvr,Milhano:2017nzm}, and \cite{Andrews:2018jcm} for a recent review on jet substructure observables in heavy-ion collisions). Despite their usefulness, a common limitation of many methods developed for \pp is that they are based on the assumption that QCD emissions are angular ordered, a property that holds in vacuum (see, e.g.\cite{Dokshitzer:1991wu}). In the presence of a medium, the phase space opens up to allow anti-angular ordered emissions\cite{CasalderreySolana:2012ef,CasalderreySolana:2011rz}. It is thus natural to expect that reclustering algorithms different from \ca could be more suitable for jet quenching studies. 

In this work, we explore a range of generalised \kt jet algorithms, with a particular focus on the formation time algorithm, hereon denoted as \ta algorithm, obtained by setting $p = 0.5$. In this case, Eq.\eqref{eq:dij} reduces approximately to
\beq
d_{ij} \approx p_{Ti} \, \theta^2 \sim \frac{1}{\tau_{form}} \, ,
\label{eq:dijtau}
\eeq
where $\theta \approx \Delta R_{ij}/R$. The formation time of an emission, $\tau_{form}$, is the time it takes to behave as an independent source of radiation, and is given by \cite{Dokshitzer:1991wu}:
\beq
\tau_{form} \approx \frac{E}{Q^2} \approx \frac{1}{2 \, E\, z (1-z) (1 - \cos \theta_{12}) } \, .
\label{eq:tau}
\eeq
The energy and virtuality of the incoming parent parton are denoted by $E$ and $Q^2$, the fraction of energy transported by the extra radiation by $z$, and $\theta_{12}$ is the emission angle between the two outgoing partons. The rightmost expression of Eq.\eqref{eq:tau} is only valid in the high energy limit. If we focus further on the collinear and soft limits, we obtain the parametric inverse of Eq.\eqref{eq:dijtau}:
\beq
\tau_{form} \approx \frac{1}{\omega \, \theta^2} \, .
\label{eq:tau_param}
\eeq

This manuscript is organised as follows: after general considerations on how jets are generated in Monte Carlo event generators (Section~\ref{sec:reco}), we  demonstrate that unclustering with $p=0.5$ can be used to extract information about formation times from reconstructed jets in \pp and PbPb collisions (Section~\ref{sec:corr}). Section~\ref{sec:hi} shows an example of an application of the proposed \ta algorithm in heavy-ion studies. The final conclusions appear in Section~\ref{sec:conclusions}. Several details and further comparisons between the different reclustering algorithms are not included in the main manuscript, but can be consulted in the appendices (A to C).

\section{Monte Carlo event generators}
\label{sec:reco}

\subsection{Jets in Monte Carlo event generators}

In the absence of a background medium, the scale evolution of jets is calculable in perturbation theory and included in Monte Carlo event generators in the form of parton showers. These generate the additional radiation through an iterative procedure, where the emissions are strictly ordered in the evolution variable. Parton showers are typically implemented in the leading logarithmic approximation. To this accuracy, the evolution variable is not uniquely determined, and a whole class of variables is allowed. The most commonly used are transverse momentum (used for instance, by \mbox{PYTHIA\,8}~\cite{Sjostrand:2006za} and SHERPA~\cite{Bothmann:2019yzt}), angle (HERWIG~\cite{Bellm:2015jjp}) and virtuality (\mbox{PYTHIA\,6}~\cite{Sjostrand:2007gs} and \jwl~\cite{Zapp:2013vla}). Inverse formation time also belongs to the class of possible evolution variables but is not widely used. A consequence of ordering in a variable other than (inverse) formation time is that the emissions are not necessarily ordered in time. In a virtuality ordered shower, like JEWEL, both energy and virtuality in Eq.\eqref{eq:tau} decrease in each emission step, but the ratio does not necessarily decrease as well. For typical kinematics, however, the occurrence of emissions un-ordered in time is rare. This justifies the common practice in jet quenching studies of using a transverse momentum or virtuality ordered parton shower with the assumption that the emissions are, in addition, ordered in time. This is the case of JEWEL, a medium modified parton shower that includes the concept of formation time to perform a veto on subsequent emissions.

\subsection{Event samples}
\label{sec:samples}

To understand how a change in the $p$ parameter can be more sensitive to in-medium effects of subsequent QCD emissions, we will use two Monte Carlo event generators: \pythia (v8.2.35, tune 4C~\cite{Corke:2010yf}) to check how these jet algorithms correlate to the vacuum shower, and JEWEL, our reference for jet quenching effects. Anti-angular ordered emissions is a feature still missing in most of jet quenching Monte Carlo event generators. In JEWEL, for instance, emissions inside the medium can populate the full accessible phase space, i.e.\, they are neither angular nor anti-angular ordered. As a first case study to identify the correlation between the emission pattern and the reclustering history, we find \jwl to be the best choice. We used an unofficial version of \jwl, based on v2.2.0, that provides the output of the parton shower history. Both generators were set to produce dijet events at a centre-of-mass energy $\sqrt{s_{NN}} = 5.02$~TeV. For the medium-induced effects, we restricted ourselves to the medium toy model provided within this event generator (Bjorken model \cite{Bjorken:1982qr} for a boost-invariant longitudinal expansion of an ideal quark-gluon gas), with an initialisation time set to $\tau_i = 0.4$~fm/c and initial temperature $T_i = 0.44$~GeV. These values are known to provide an $R_{AA} \simeq 0.4$, in agreement with current experimental observations at the same centre-of-mass energy  \cite{Acharya:2018qsh,Aaboud:2018twu,CMS:2019btm} (see also section \ref{sec:hi}). Two PbPb event samples are used: one with recoils from elastic scattering to include the effects of medium response, and one where these particles are discarded from the final event. When recoils are included, the subtraction of thermal momenta is performed using the new constituent subtraction algorithm~\cite{paperKorinnaNew}, which is particularly suitable for jet substructure studies. Unless noted otherwise, the JEWEL results shown are without recoils, so as not to complicate the discussion.

\section{Correlation between parton shower and jet history}
\label{sec:corr}

\subsection{Vacuum parton shower}
\label{sec:corr_vac}

From hadronic \pythia dijet events we reconstruct \akt jets with $R= 0.5$. After identifying the leading jet with a minimum transverse momentum $p_{T,min} = 300$~GeV that falls within $|\eta_{jet}| < 1.0$, we recluster the jet particles with the generalised \kt algorithm, for several $p$ values (setting $R = 1.0$ for the reclustering). These steps are performed within FastJet v3.3.0\cite{Cacciari:2011ma}. 

We then iteratively uncluster the reclustered jet, following the leading branch. At each step, we evaluate the formation time as provided by the rightmost expression in Eq.\eqref{eq:tau}, using the information from the two obtained subjets\footnote{Note that the angle $\theta_{12}$ is evaluated as the angle between the three-dimensional momentum vectors of the two subjets.}. This procedure yields the unclustering history. The obtained sequence is not necessarily ordered in formation time, when a distance measure different from (inverse) formation time is used for the reclustering. In practice, however, the first unclustering step will usually be the one with the shortest formation time. The probability that this is not the case was found to be a few percent in this study (as expected, it is larger for \ca than for the \ta algorithm\footnote{It is not zero even with the \ta algorithm because its distance measure equals the inverse of Eq.\eqref{eq:tau} only in the soft and collinear approximation.}. Therefore, we always assume that the first unclustering step has the shortest formation time (instead of walking through the whole sequence to find the step with the shortest formation time).

To correlate the unclustering sequence with the parton shower, we identify the initial parton (coming directly from the matrix element) that generated this jet\footnote{To overcome the difficulties in following the parton shower history when hadronisation effects are taken into account, we identify all the final state particles produced by the two ancestors. We reclustered them with \akt jet algorithm and $R = 1$. The ancestor that generated the jet is considered to be the one whose jet axis is within $\Delta R < 1.0$. For more information, we refer the reader to \ref{app:Details}.} and, following the leading branch, calculate in each step the splitting kinematics as the rightmost expression in Eq.~\eqref{eq:tau}. We also studied the impact of applying the high-energy limit, for which we refer the reader to \ref{app:tau12}.

\begin{figure}[h!]
\centering
\subfloat[Reclustering with \ca algorithm]{
\centering
\includegraphics[width=0.9\columnwidth]{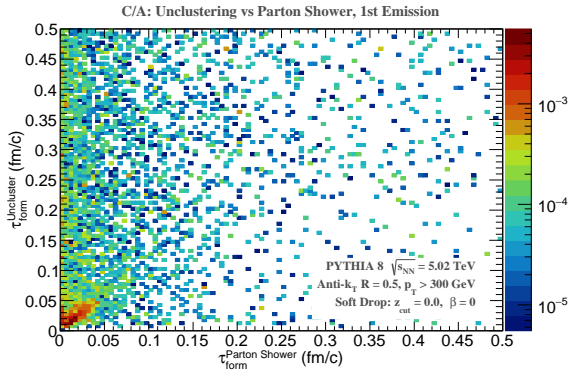}
\label{fig:CorrCA}}
\linebreak
\subfloat[Reclustering with \ta algorithm]{
\centering
\includegraphics[width=.9\columnwidth]{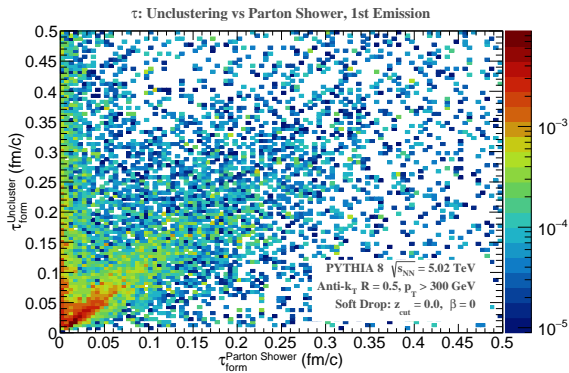}
\label{fig:CorrTau}}
\caption{Correlation of \taf between the first parton shower emission and the first unclustering step for different generalised \kt algorithms in \pythia.}
\label{fig:Corr}
\end{figure}

The correlation of the obtained \taf between the first unclustering step and the first parton shower emission in \pythia is shown in Fig.~\ref{fig:Corr}. The top (Fig.~\ref{fig:CorrCA}) and bottom (Fig.~\ref{fig:CorrTau}) panels are for jets reclustered with the \ca ($p = 0$) and \ta algorithm ($p = 0.5$), respectively. Two features are present: (i) the diagonal elements that represent the events in which the correlation is positive (true) (ii) the vertical elements that correspond to large angle emissions that are not captured by the final jet, which leads to a mismatch between the first emission of the parton shower and the first emission captured by the reconstructed jet. Naturally, this vertical component is reduced if we increase the jet radius. 

Without any jet grooming technique, the correlation between the unclustering and the parton shower is more pronounced for the \ta algorithm. However, this is not an entirely fair comparison, as the \ca algorithm is particularly sensitive to uncorrelated soft large-angle emissions, but is able to provide a proxy for the Altarelli-Parisi splitting function if a SoftDrop procedure is used. Following~\cite{Larkoski:2014wba}, we apply such a condition to select only emissions (during the parton shower) and/or unclustering steps (during the jet unclustering process) with:
\beq
z_g = \frac{ \min(p_{T1},p_{T2}) }{ p_{T1}+p_{T2} } > z_{cut} \left( \frac{\Delta R_{12} }{R_{jet}} \right)^{\beta} \, .
\label{eq:softdrop}
\eeq

\begin{figure}[h!]
\centering
\subfloat[Reclustering with \ca algorithm]{
\centering
\includegraphics[width=0.9\columnwidth]{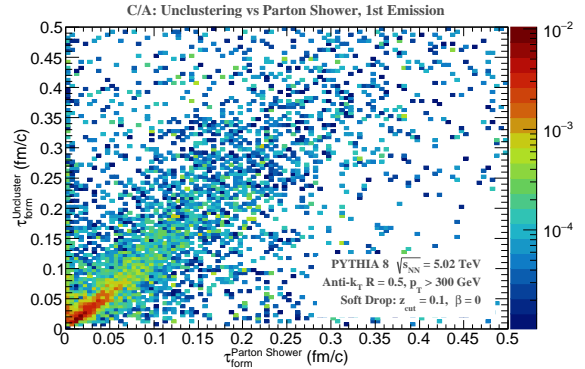}
\label{fig:CorrCA2}}
\linebreak
\subfloat[Reclustering with \ta algorithm]{
\centering
\includegraphics[width=.9\columnwidth]{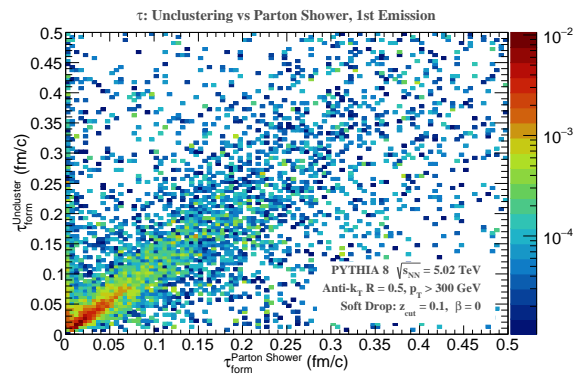}
\label{fig:CorrTau2}}
\caption{Correlation of \taf between the first parton shower emission and the first unclustering step for different generalized \kt algorithms with SoftDrop grooming ($z_{cut} = 0.1$ and $\beta = 0$) in \pythia.}
\label{fig:Corr2}
\end{figure}

Fig.~\ref{fig:Corr2} shows the correlation between \taf when a SoftDrop procedure with  $z_{cut} = 0.1$ and $\beta = 0$ is used. The top panel illustrates the correlation when the jet reclustering is done with \ca, and the bottom panel for \ta algorithm. 

Compared with the results without SoftDrop (Fig.~\ref{fig:Corr}), there is a significant decrease of the dispersion in both cases. The vertical events aligned at $\tau_{form}^{Parton Shower} \simeq 0$ are also drastically reduced. These correspond to relatively soft emissions during the parton shower that would fall outside of the jet cone. With the introduction of a \zcut, these emissions are discarded and the match between parton shower and unclustering history improves. Overall, the correlation factor increased from 0.26 (C/A) and 0.38 ($\tau$) to 0.65 (C/A) and 0.66 ($\tau$). As such, the use of a SoftDrop procedure is recommended in order to increase the correlation between the jet clustering history and the parton shower. 

\begin{figure}[h!]
\centering
\includegraphics[width=.9\columnwidth]{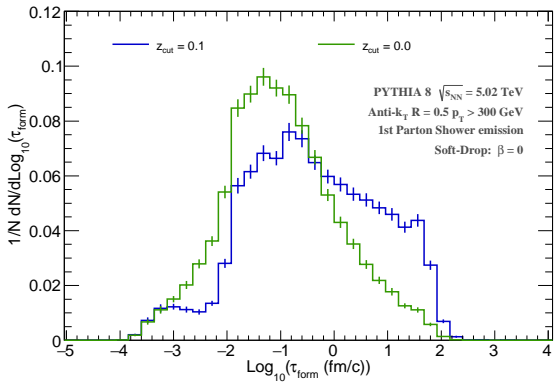}
\caption{Distribution of $\log_{10}(\tau_{form})$ for the first emission obtained from \pythia parton shower when using $z_{cut} = 0.1$ (blue) and $z_{cut} = 0.$ (green).}
\label{fig:time_z0}
\end{figure}

The effect of SoftDrop grooming is also reflected in the overall \taf distribution. In Fig.~\ref{fig:time_z0}, we show $\log(\tau_{form})$ of the first emission as obtained from the \pythia parton shower. The blue line represents the distribution when grooming is introduced, and the green when $z_{cut} = 0.0$. Without grooming, the \taf distribution is dominated by (early) large angle and soft emission. With $z_{cut} = 0.1$, these emissions are discarded and, consequently, the selected first emission has a significantly larger \taf.
In the following, we will keep $z_{cut} = 0.1$ fixed, unless noted otherwise.

\begin{figure}[h!]
\centering
\includegraphics[width=.9\columnwidth]{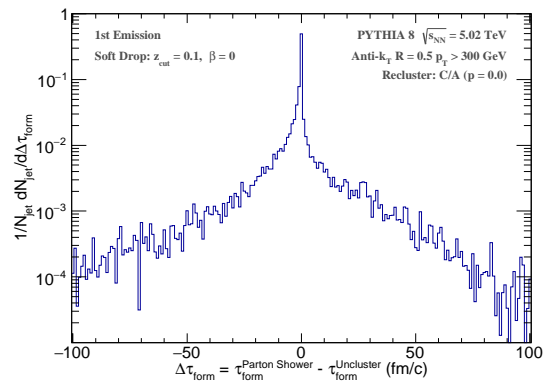}
\caption{$\Delta \tau$ distribution as defined in Eq.~\eqref{eq:DeltaTau}, for the first emission/unclustering step with the \ca jet algorithm ($p = 0.0$) and $z_{cut} = 0.1$.}
\label{fig:deltaTau}
\end{figure}

\begin{figure}[h!]
\centering
\includegraphics[width=\columnwidth]{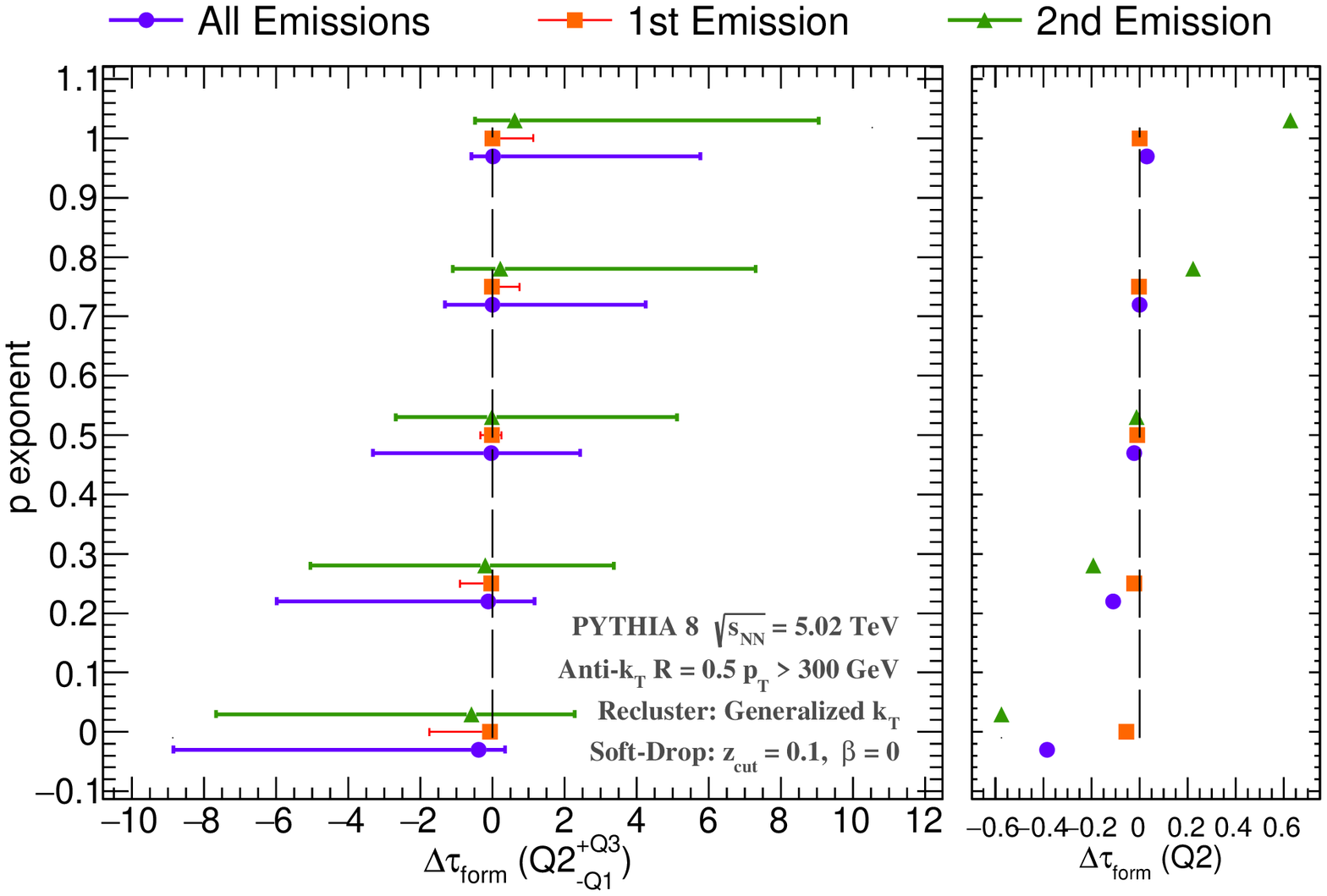}
\caption{Median value of the $\Delta \tau$ distribution (Eq.~\eqref{eq:DeltaTau}) obtained for different $p$ parameters defined in the generalised \kt jet algorithms for the first (orange), second (green) or all emissions along the primary branch (purple) in \pythia events. The asymmetric error bars correspond to $+Q_3$ and $-Q_1$ quartiles. The right panel shows a zoom of $Q_2$ alone.}
\label{fig:evolp}
\end{figure}

To quantify the correlation and compare with other values of $p$, we define the difference of the two \taf defined in Figs.~\ref{fig:Corr} and \ref{fig:Corr2}, $\Delta \tau$, for each event and each splitting, as
\beq
\Delta \tau = \tau_{form}^{Parton \ Shower} - \tau_{form}^{Unclustering} \, .
\label{eq:DeltaTau}
\eeq
If the correlation is successful, the resulting distribution should be narrow and peaked at zero. As an illustration, we show, in Fig.~\ref{fig:deltaTau}, the distribution for the difference defined by Eq.~\eqref{eq:DeltaTau} obtained for the first emission/unclustering step with the \ca algorithm.
From these distributions, we calculate the median $Q_2$, and the $Q_1$ and $Q_3$ quartiles as a measure of the width of distribution. These results are collected in Fig.~\ref{fig:evolp} from \pythia for different $p$ values. In the left panel, we show the median (marker) and, since the distribution is generally asymmetric, we keep $-Q_1$ and $+Q_3$ in the form of asymmetric error bars. We also consider separately the first splitting (in orange), the second (in green), and all the primary branch (in purple), slightly offset from the central values ($p = 0.0; 0.25; 0.5; 0.75; 1.0$) to improve readability. In the right panel, we show a zoom for $Q_2$ alone.
Generally, all $p$ parameters yield median values close to 0 (right panel of Fig.~\ref{fig:evolp}.), in particular for the first emission (in orange). However, only central $p$ values in the considered range ($p\in[0,1]$) yield a symmetric and narrow distribution. As we proceed further through the primary parton shower branch/unclustering history, we start to see deviations in the median value. In particular, the second emission only has a peaked distribution around $0$ for $p\simeq 0.5$. It is also substantially broader, even with SoftDrop grooming. These effects are also visible when considering all emissions along the primary branch. However, we do not see an increasing effect with respect to the second emission, as this distribution is dominated by the first emission (many jets consist of only two subjets after grooming and thus have only one unclustering step). In addition, the average formation time also increases for the second splitting, naturally leading to a larger interquartile range\footnote{For each splitting,  the $\Delta \tau_{form}$ distribution contains all timescales. Since large values of $\tau_{form}$ tend to have larger values of $\Delta \tau_{form}$, the IQR increases with $\tau_{form}$. But the $\tau_{form}$ distribution is dominated by small values of $\tau_{form}$. The IQR extracted from the inclusive $\Delta \tau_{form}$ distribution does therefore not represent an average resolution.} (IQR $= Q_3 - Q_1$). When comparing the relative differences instead of absolute differences the resolution for the second splitting is identical to the first splitting. Nonetheless, in this manuscript, we will restrict ourselves to illustrate an application of using the first unclustering step as a tool for jet quenching studies. 

\begin{figure}[h!]
\centering
\includegraphics[width=\columnwidth]{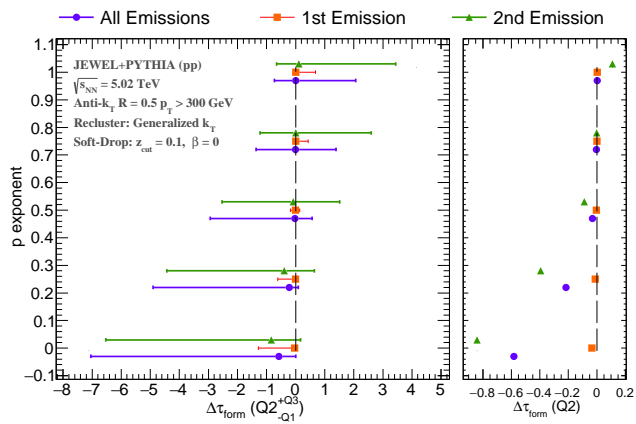}
\caption{Median value of the $\Delta \tau$ distribution (Eq.~\eqref{eq:DeltaTau}) obtained for different $p$ parameters defined in the generalised \kt jet algorithms for the first (orange), second (green) or all emissions along the primary branch (purple) in JEWEL+PYTHIA (pp) events. The asymmetric error bars correspond to $+Q_3$ and $-Q_1$ quartiles. The right panel shows a zoom of $Q_2$ alone.}
\label{fig:JewelVac}
\end{figure}

To understand if the apparent success of the correlation of the \ta algorithm with the parton shower history is not dominated by details of the Monte Carlo event generator, we repeated the exercise in \jwl without medium effects (see Fig.~\ref{fig:JewelVac}).
The magnitude of the observed shifts and dispersions is similar to the ones obtained in \pythia. The main differences are observed for large values of $p$, where the shift in the distribution of the second emission is more compatible with zero. While the results differ in details, the conclusions are still qualitatively the same: intermediate values of $p \in [0,1]$ yield a better correlation between jet algorithm and parton shower history, with the \ta algorithm ($p=0.5$) showing the best performance. In particular, it has the most symmetric distribution, which is important for obtaining an unbiased estimate of the formation time. Such an outcome was expected as the \ta algorithm uses (inverse) formation as distance measure in the clustering. The fact that the findings are so similar for two rather different parton showers (in particular with different ordering variables) gives us confidence in the robustness of the procedure and why thus turn to study its performance in central PbPb collisions.

\subsection{Medium-modified parton shower}
\label{sec:corr_jwl}

\begin{figure}[h!]
\centering
\includegraphics[width=.9\columnwidth]{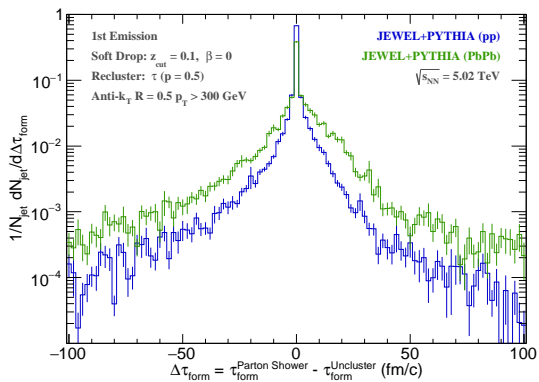}
\caption{$\Delta \tau$ distribution as defined in Eq.~\eqref{eq:DeltaTau}, for the first emission/unclustering step with \ta jet algorithm ($p = 0.5$), and $z_{cut} = 0.1$. The results in blue refer to JEWEL+PYTHIA (pp), and in green to JEWEL+PYTHIA (PbPb.)}
\label{fig:deltaTauJWL}
\end{figure}

We now include jet quenching effects, and we turn our attention to the results provided by \jwl with the medium model settings as described in section \ref{sec:samples}. To illustrate the change in the $\Delta\tau$ distribution, we show, in Fig.~\ref{fig:deltaTauJWL}, the results for JEWEL+PYTHIA (pp) (in blue) and JEWEL+PYTHIA (PbPb) (in green), for the first emission\footnote{We ignore the elastic interactions vertices.}/unclustering step obtained with the \ta algorithm.
An immediate consequence of including medium effects is the overall smearing of the $\Delta \tau$ distribution. This is due to further interactions of the partons coming from the first splitting in the medium. They lead to a deterioration of the correlation between the partons' and the subjets' momenta, and limit the feasibility of reconstructing the kinematics at the splitting vertex from the final state particle distribution. Energy loss (energy transported to large distances) and transverse momentum broadening are thus a limiting factor in the endeavour to estimate the formation time of the first splitting in the presence of a background medium (this comes on top of effects like vacuum-like emissions and contamination from initial state radiation governing the resolution in pp). In addition, the average formation time increases in PbPb compared to pp (see discussion in section~\ref{sec:hi}), and the correlation between reconstructed and true formation time deteriorates for larger \taf (cf. fig.~\ref{fig:Corr2}).

\begin{figure}[h!]
\centering
\includegraphics[width=\columnwidth]{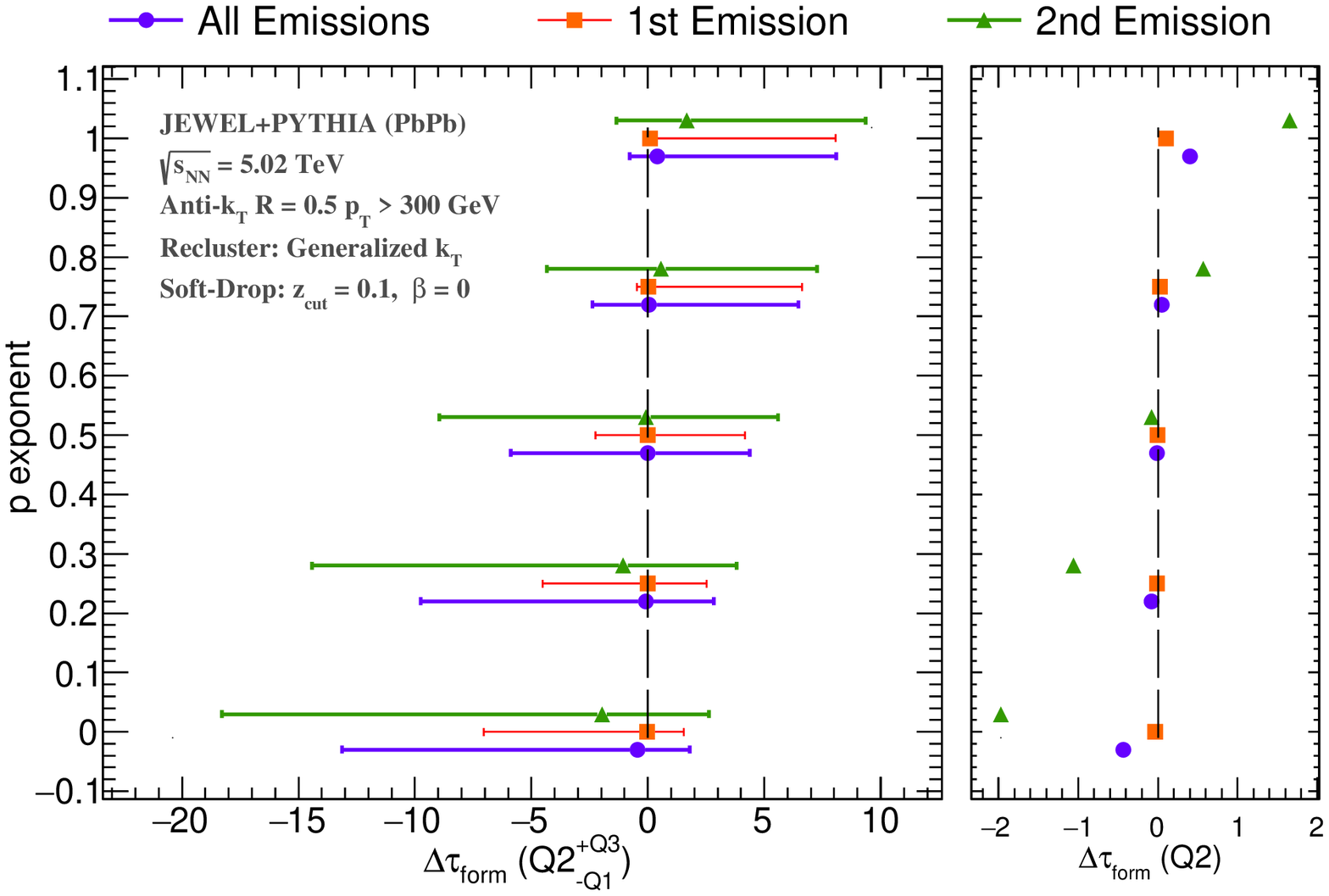}
\caption{Median value of the $\Delta \tau$ distribution (Eq.~\eqref{eq:DeltaTau}) obtained for different $p$ parameters defined in the generalised \kt jet algorithms for the first (orange), second (green) or all emissions along the primary branch (purple) for \jwl (PbPb) events. The asymmetric error bars correspond to $+Q_3$ and $-Q_1$ quartiles. On the right, a zoom of the $Q_2$ alone is shown.}
\label{fig:evolpJWL}
\end{figure}

The characteristics of the $\Delta \tau$ distributions obtained from the unclustering history and the parton emission history in \jwl (PbPb) for different exponents of the generalized \kt algorithm are shown in Fig.~\ref{fig:evolpJWL}. 
As expected, the smearing observed in Fig.~\ref{fig:deltaTauJWL} is also present for subsequent emissions and across all $p$ exponents.  Nonetheless, the same qualitative features appear as in Fig.~\ref{fig:evolp}: $p\simeq0.5$ provides the best correlation (less biased and narrower $\Delta \tau$) between parton shower and jet unclustering history. 

\begin{figure}[h!]
\centering
\includegraphics[width=\columnwidth]{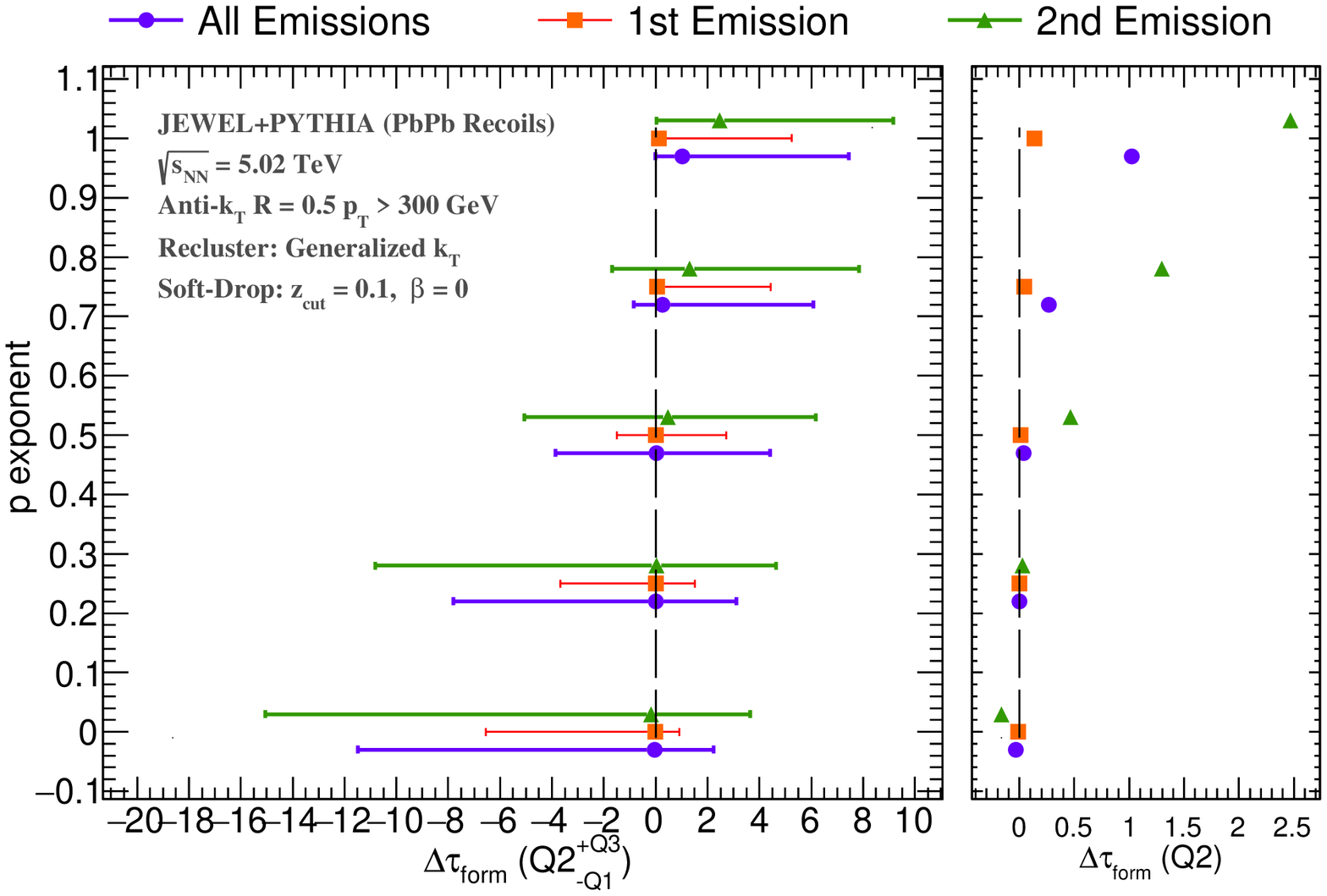}
\caption{Median value of the $\Delta \tau$ distribution (Eq.~\eqref{eq:DeltaTau}) obtained for different $p$ parameters defined in the generalised \kt jet algorithms for the first (orange), second (green) or all emissions along the primary branch (purple) for \jwl (PbPb) events with recoils. The asymmetric error bars correspond to $+Q_3$ and $-Q_1$ quartiles. On the right, a zoom of the $Q_2$ alone is shown.}
\label{fig:evolpJWLRec}
\end{figure}

When recoils are considered, part of the energy lost by the jet is recovered (in the form of additional soft activity inside the jet cone). Including medium response thus mitigates the mismatch between subjet kinematics and kinematics at the splitting vertex. At the same time, it also introduces as additional smearing: as the additional activity from medium response has a broad distribution, subjets will, in general, also include some level of activity from other sources than its parent parton. Overall, there remains a net improvement of the correlation reducing the width of the distribution, independently of the $p$ exponent in the generalized \kt algorithms (see Fig.~\ref{fig:evolpJWLRec}). This effect is more noticeable in the second emission: being softer, the effect of the medium recoils is, proportionally, larger than for the first unclustering step. Given that the nature of medium recoils is still under debate, we will focus on \jwl without considering this medium recoil component, thus simplifying following discussions.

We also investigated the role of the jet radius and Soft-Drop $z_{cut}$ parameter, with focus on \ca and \ta clustering algorithms (see \ref{app:evolSettings} and \ref{app:tau12} for the corresponding results without the high-energy approximation). As expected, an increase of the jet radius increases the correlation and reduces the smearing of the $\Delta \tau$ distribution, for both first and second emission. Large radius jets, however, pose a challenge in a heavy-ion environment due to the high event activity~\cite{CMS:2019btm,ATLAS:2019rmd}. As for the effect of varying $z_{cut}$, there is a trade-off between reducing contamination from the uncorrelated soft activity and discarding correlated structures. Overall, we found that $z_{cut} \simeq 0.1$ seems to be a good compromise.

\subsection{Summary}
\label{sec:CorrelationSummary} 

To present a summary of the main properties of the $\Delta \tau$ distribution, we focus on the first and second emissions of the leading \akt jet with $R = 0.5$ and $p_{T} > 300$~GeV, reclustered with the generalized \kt algorithm and Soft-Drop procedure ($z_{cut} = 0.1$, $\beta = 0$). 

\begin{figure}[h]
\centering
\includegraphics[width=\columnwidth]{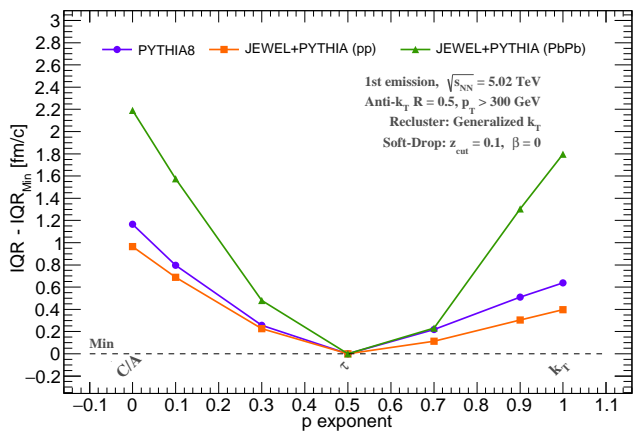}
\caption{Difference of interquartile range (IQR = $Q_3 - Q_1$) of the $\Delta \tau$ distribution (Eq.~\eqref{eq:DeltaTau}) obtained for different $p$ parameters with respect to the minimum value. The results when using \pythia are shown in purple, \jwl (pp) in orange, and \jwl (PbPb) in green.}
\label{fig:Summary1}
\end{figure}

The median of the distribution for the first emission is compatible with $0$ and so we only focus on the interquartile range (IQR $= Q_3 - Q_1$). To provide a better visualisation given the different magnitudes between pp and PbPb, we identify the minimum IQR value obtained from the different $p$ exponents used in the generalized \kt algorithm and subtract this amount, such that the minimum is zero for each of different samples. These are shown in Fig.~\ref{fig:Summary1} for \pythia(purple/circle markers), \jwl (pp) (orange/square markers) and \jwl (PbPb) (green/triangle markers). All models consistently indicate that $p=0.5$ yields the smallest dispersion. When \ca is used instead, the width of the distribution increases by an additional $1$~fm/c (pp) up to $2.2$~fm/c (PbPb).

\begin{figure}[h]
\centering
\includegraphics[width=\columnwidth]{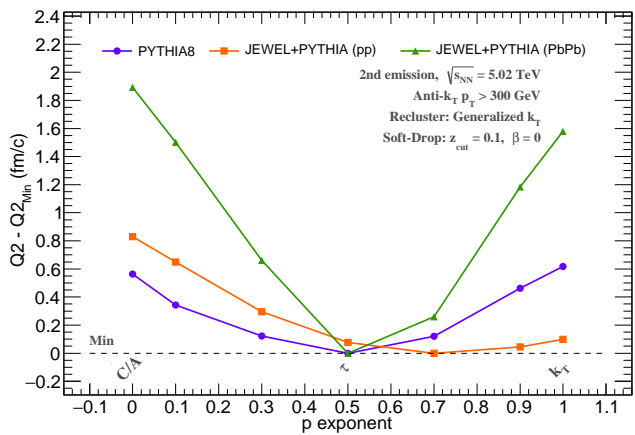}
\caption{Difference of median ($Q_2$) of the $\Delta \tau$ distribution (Eq.~\eqref{eq:DeltaTau}) obtained for different $p$ parameters with respect to the minimum value. The results when using \pythia are shown in purple, \jwl (pp) in orange and \jwl (PbPb) in green.}
\label{fig:Summary2}
\end{figure}

For the distribution of the second emission we focus on the median, as there are significant differences between the clustering algorithms. Again, we take the difference with respect to the minimum value obtained for the different $p$ exponents. The results for the different Monte Carlo parton showers are shown in Fig.~\ref{fig:Summary2}. In this case, $p = 0.5$ seems to be preferred by \jwl (PbPb), but vacuum parton showers do not agree on the exponent that maximises the correlation: with \pythia intermediate values yield a better correlation, while \linebreak \jwl (pp), which is based on PYTHIA 6, shows a less biased distribution when larger values of $p$ are used instead. These differences might be related to the details of the implementation of the parton shower, including small violations to the time ordering of the emissions. Nonetheless, the differences between the two vacuum parton showers are up to $0.5$~fm/c for the \kt algorithm. As remarked earlier, we here focus on the first splitting.

These results suggest that the use of unclustering tools others then standard \ca for jet quenching studies might potentiate an accurate evaluation of the QGP properties. In particular, the use of the \ta algorithm for reclustering allows to directly extract an estimate of the formation times in a more meaningful way. The use of formation time from splitting processes is already under exploration by the community~\cite{Apolinario:2020zvt,Citron:2018lsq}. It will, for sure, bring a novel approach to probe the QGP evolution, following recent efforts~\cite{Apolinario:2017sob,Andres:2019eus}. By using the \ta algorithm for reclustering, we can increase the precision of such studies and access more differential QGP timescales. While such a feasibility study will be left for a future communication, we will proceed with an example to compare the use of \ta and \ca for jet quenching studies.

\section{Comparison of \ta and \ca reclustering algorithms for jet quenching studies}
\label{sec:hi}

\begin{figure}[h!]
\centering
\includegraphics[width=.9\columnwidth]{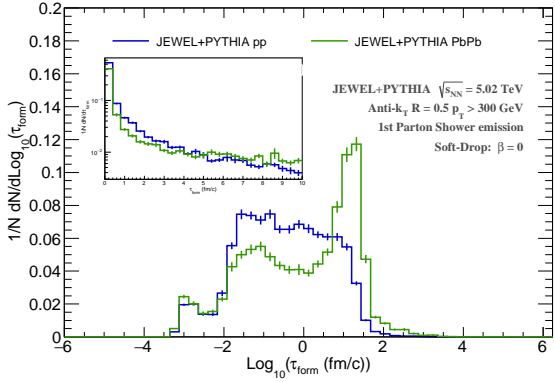}
\caption{$\tau_{form}$ distribution for the first parton shower emission obtained from \jwl in pp (blue) and PbPb (green) when using $z_{cut} = 0.1$. The distribution is shown as a function of $log_{10}(\tau_{form})$, and the inset shows the distribution on linear-log scale.}
\label{fig:tauform_mc}
\end{figure}

In Fig.~\ref{fig:tauform_mc}, we compare the formation time of the first groomed emission obtained in \jwl pp (blue) and PbPb (green) from the parton shower. There is a clear increase in the average formation time in PbPb with respect to pp. Several effects contribute to a change of the resulting \taf distribution when medium-induced effects are considered. The most immediate consequence -- energy loss (both elastic and inelastic) -- leads to a decrease of \taf. In the case of elastic energy loss this occurs through the degradation of the energy in eq.~(\ref{eq:tau}) (the virtuality does not change), while in the case of inelastic energy loss it happens via the occurrence of an extra splitting (with formation time shorter than the vacuum-like splitting). However, both effects are small for the first emission. Elastic energy loss is not important since the parton is very energetic, and inelastic scattering is extremely rare due to the very short formation time of the first vacuum-like splitting. On top of energy loss, QGP interactions also induce jet collimation, an effect that has already been observed in other observables~\cite{CasalderreySolana:2010eh,Apolinario:2017qay,Andrews:2018jcm}. The surviving jets in PbPb collisions are biased towards having a hard (vacuum-like) fragmentation pattern, which is correlated to larger formation times. This effect is dominant for the first emission and contributes to the overall shift of the formation time distribution observed in Fig.~\ref{fig:tauform_mc}.

\begin{figure}[h!]
\centering
\includegraphics[width=\columnwidth]{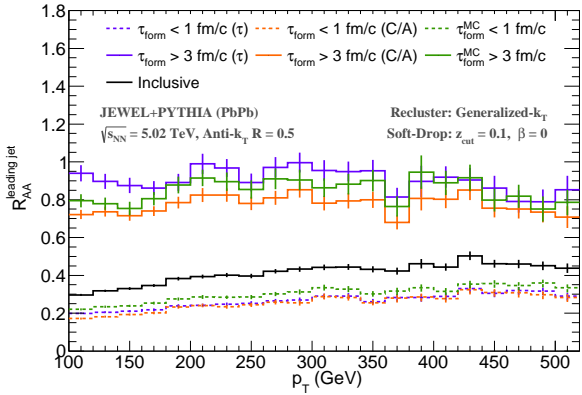}
\caption{\jwl nuclear modification factor of leading jets. The jets reclustered with \ta (C/A) algorithm are shown in purple (orange) when selecting the first groomed unclustering step with  $\tau_{form} > 3$~fm/c ($\tau_{form} < 1$~fm/c) in solid (dashed) lines. For reference, we add the results when reading the \taf from the Monte Carlo parton shower in green, and the inclusive spectrum in solid black.}
\label{fig:raa_tau1_worec}
\end{figure}

We now proceed to evaluate the nuclear modification factor of leading jets provided by JEWEL, but extending this study to lower transverse momentum ($p_{T} > 100$~GeV). Using \ca and \ta to recluster the leading jet, we identify the first unclustering step and  create two populations: \emph{early jets}, whose first unclustering step has $\tau_{form} < 1$~fm/c, and \emph{late jets}, with a first unclustering step with $\tau_{form} > 3$~fm/c. We make this selection in \jwl (PbPb) and \jwl (pp) events to obtain the leading jet transverse momentum spectrum in both cases. The corresponding medium-over-vacuum ratio (nuclear modification factor, $R_{AA}$, for leading jets) is shown in Fig. ~\ref{fig:raa_tau1_worec}. For reference, we also include the inclusive leading jet ratio, in solid back. The purple lines refer to reclustering with \ta algorithm for late (solid line) and early (dashed line) jets, while the orange refers to the \ca algorithm. For reference, we also include the results directly read from the parton shower, in green.
There is a clear difference in the leading jet suppression when, instead of using the full sample, we select jets whose fragmentation starts shortly after its production. These jets are, as expected, strongly suppressed, and both \ca, and \ta algorithms provide similar results. Taking the results from section \ref{sec:corr}, we do not expect to see much deviations between the two. However, as we move towards late times, the two algorithms show some differences. In particular, if we use \ta to recluster the jet particles, the obtained $R_{AA}$ is compatible with $1$. As discussed earlier, these jets have a hard fragmentation pattern and are therefore not so susceptible to modifications due to medium interactions as those with a soft fragmentation, and thus early first splitting. In particular, the late jets consist of only one effective colour charge with high momentum (in this case $\sim \unit[300]{GeV}$) for the first \unit[3]{fm} of the evolution. This object loses little energy through elastic scattering, and, when it finally splits, the medium density is already diluted ($\epsilon < 5$~GeV/fm$^3$ for the medium settings used here and the simple medium model). At relatively low $p_{T}$, both algorithms yield a similar difference with respect to the  Monte Carlo truth (one suppressed, the other enhanced), but at high \pt, the results using the \ta algorithm approaches the Monte Carlo. This is in line with the observations of the previous section.

\begin{figure}[h!]
\centering
\includegraphics[width=\columnwidth]{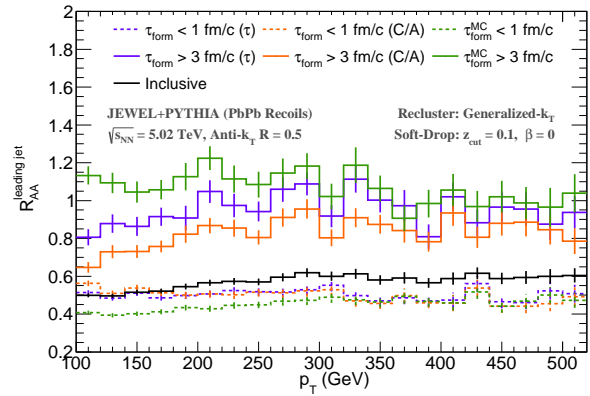}
\caption{\jwl nuclear modification factor of leading jets when recoils are considered. The jets reclustered with \ta (C/A) algorithm are shown in purple (orange) when selecting the first groomed unclustering step with  $\tau_{form} > 3$~fm/c ($\tau_{form} < 1$~fm/c) in solid (dashed) line. For reference, we added the the results when reading the \taf from the Monte Carlo parton shower in green, and the inclusive spectrum in solid black.}
\label{fig:raa_tau1_wrec}
\end{figure}

When medium recoils are considered, we see the same behaviour, see Fig.~\ref{fig:raa_tau1_wrec} (same colors and line settings as in Fig.~\ref{fig:raa_tau1_worec}).  
The \emph{early} (and inclusive) leading jet $R_{AA}$ are now slightly larger, as part of the energy is recovered by the presence of recoils. Both reclustering algorithms continue to yield the same results. However, for \emph{late} jets, there are sizable differences between the two reclustering algorithms. The difference between \ca and \ta is related to the fact that the \taf values extracted with \ca are systematically too small, particularly for large \taf (cf. figs~\ref{fig:Corr2}, \ref{fig:JewelVac} and \ref{fig:deltaTauJWL}). The result obtained with the \ta algorithm comes closer to the one extracted directly from the parton shower. This highlights the importance of an unbiased estimate of the true formation time and is a strong argument for using the \ta algorithm to obtain the formation times.

\section{Conclusions}
\label{sec:conclusions}

In this work, we explore the use of generalised \kt algorithms to correlate the parton shower history to the unclustering sequence for the primary branch. We focus on the formation time, $\tau_{form}$, to have a space-time picture of the parton shower emissions. This study is based on Monte Carlo event generators with different ordering variables (\pythia and JEWEL) for subsequent emissions. The aim was to check the resilience of our results. As exponents for the generalized \kt algorithm we focus on $p = 0$ (C/A) and $p = 0.5$ ($\tau$). We found that the formation times extracted with the \ta algorithm generally correlate better with the  parton shower values, even in the absence of SoftDrop grooming. When a $z_{cut}$ is introduced, both algorithms yield relatively small differences in \taf, but \ca shows consistently a broader and more asymmetric      distribution. Naturally, the fraction of unclustering sequences that are not ordered in formation time is larger for \ca. This could, to some extent, be overcome by going through the entire sequence to find the unclustering step with the shortest formation time (instead of always taking the first one). While in pp collisions the angular ordered sequence obtained with the \ca algorithm is often preferred due to its similarity with the QCD radiation pattern, this is not necessarily the case in heavy ion collision. As an example, this study shows that to estimate formation times it is advantageous to recluster the jets with the \ta algorithm, that interprets the fragment distribution in terms of inverse formation time. As an example, we found that the \taf obtained with \ca yields a leading jet nuclear modification factor of around 0.8 for jets whose fragmentation starts after \taf = 3~fm/c, while the \ta algorithm does not yield any suppression. This observation is consistent with the event generator results for jets with $p_T > 300$~GeV.

Time differential measurements of the QGP are increasingly needed. Jet reclustering tools that allow to increase the precision with which such short timescales can be extracted  will undeniably be most helpful in unlocking the use of jets as precise tools for QGP tomographic measurements. In the future, we plan to use these tools to perform a feasibility study in a heavy-ion environment.

\vspace{2mm}
\textbf{Acknowledgments}
    The authors would like to thank to R. Concei\c{c}\~{a}o, L. Cunqueiro, G. Milhano and J. Thaler for fruitful discussions. LA would like to thank to J. Rufino for his support with the computational resources where these analysis were performed (Research Centre in Digitalization and Intelligent Robotics  -- CeDRI -- at Instituto Politécnico de Bragança). LA acknowledges the financial support by OE - Portugal, Funda\c{c}\~{a}o para a Ci\^{e}ncia e Tecnologia (FCT) under contract \linebreak DL57/2016/CP1345/CT0004 and project CERN/FIS-PAR/0024/2019, and by the European research Council project ERC-2018-ADG-835105 YoctoLHC. The work by KZ has  received funding from the European Research Council (ERC) under the European Union’s Horizon 2020 research and innovation programme (Grant agreement No. 803183 collectiveQCD).

\appendix

\section{Identification of the jet initiating parton}
\label{app:Details}

\begin{figure}[h!]
\centering
\subfloat[Parton level]{
\centering
\includegraphics[width=.9\columnwidth]{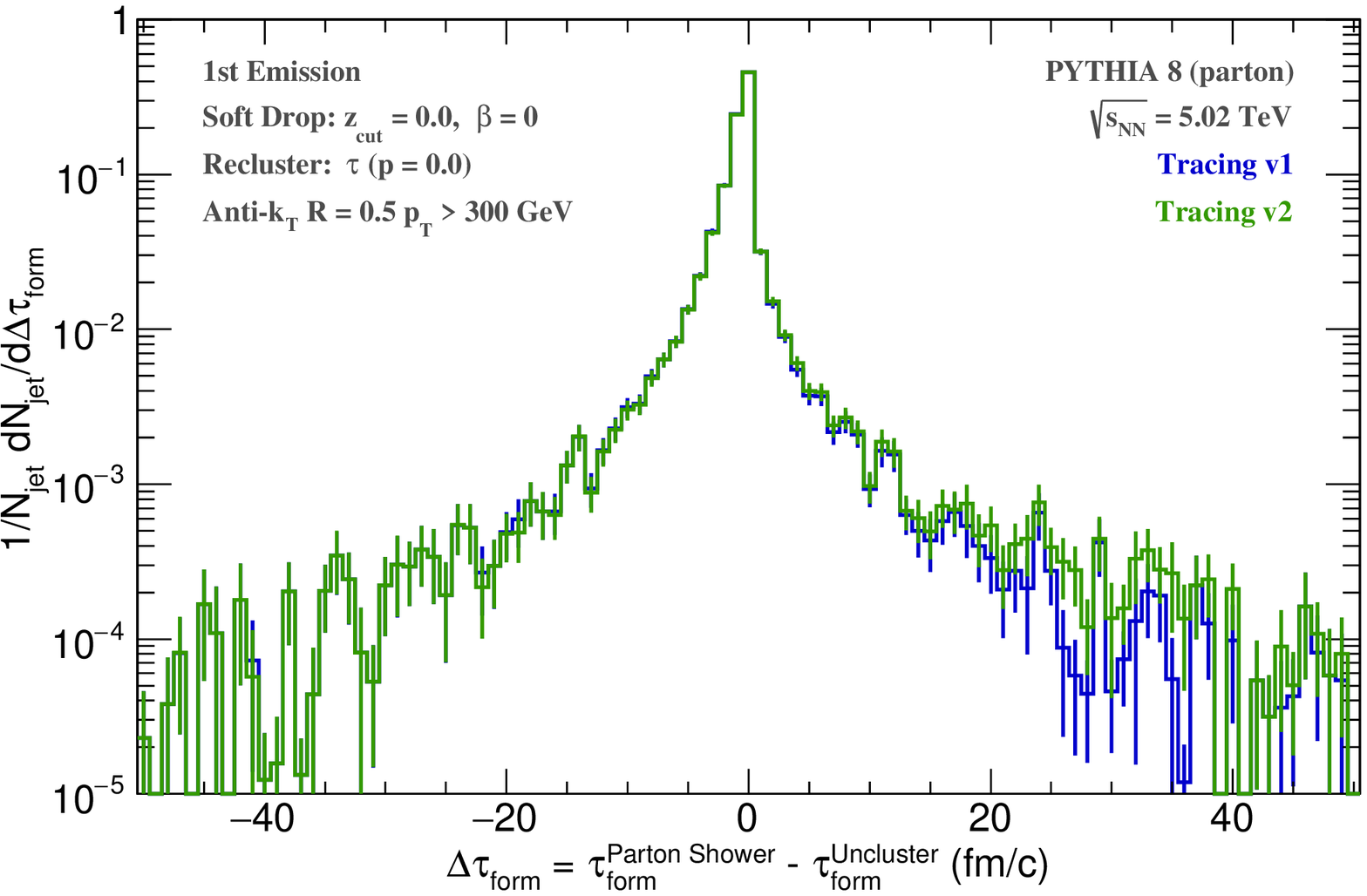}
\label{fig:compParton}}
\linebreak
\subfloat[Hadron level]{
\centering
\includegraphics[width=.9\columnwidth]{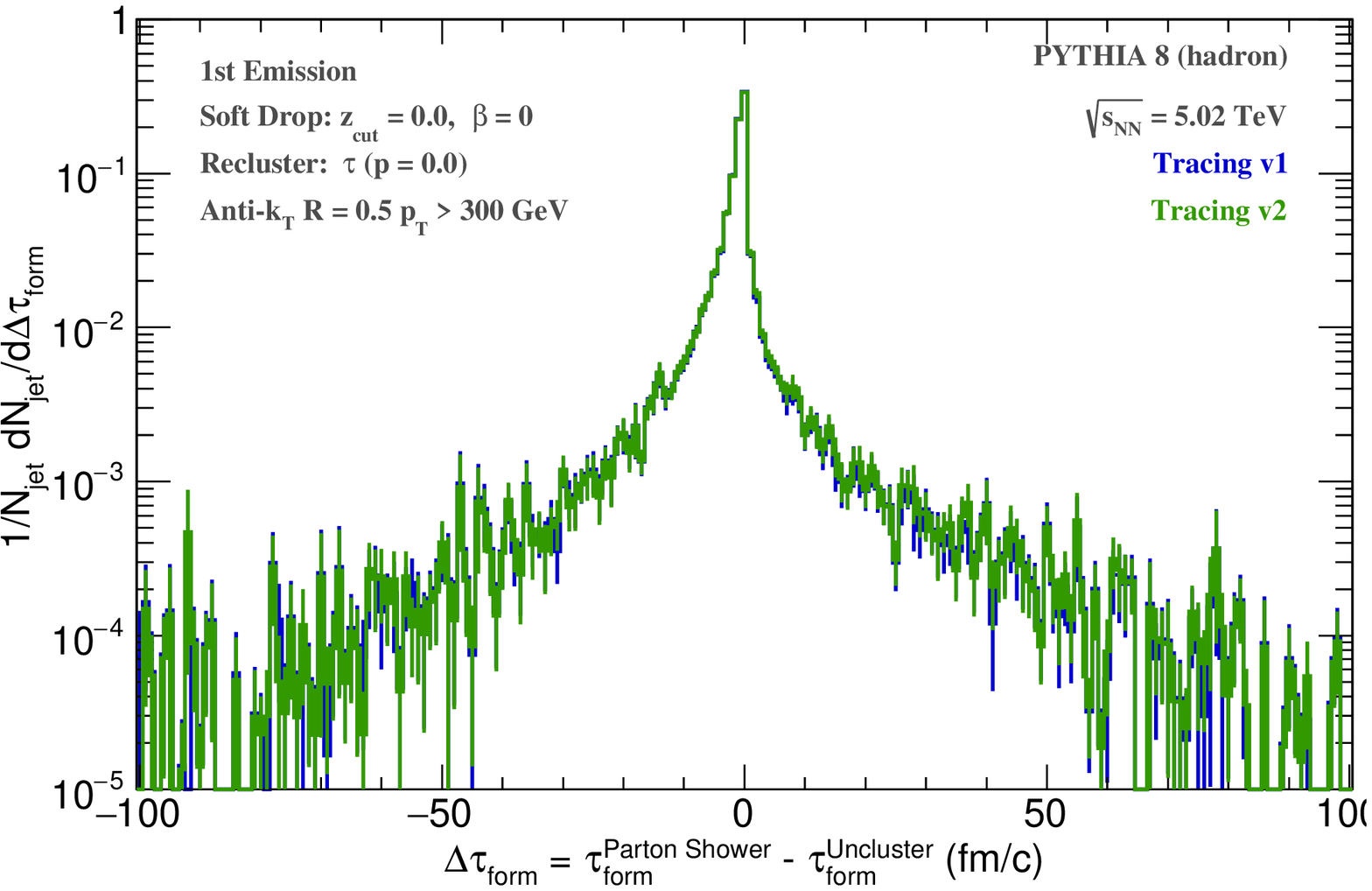}
\label{fig:compHadron}}
\caption{$\Delta\tau$ distribution for \pythia events at parton (top) and hadron level (bottom) obtained by following the two versions of the tracing procedure: v1 in blue and v2 in green (see text for an explanation of the two versions).}
\label{fig:compTime}
\end{figure}

The identification of the jet initiating particle can, in PYTHIA\,8, be done easily by looking at the ancestor list of the jet constituents (denoted as v1 in Fig.~\ref{fig:compParton} and \ref{fig:compHadron}). The initiating parton is taken to be the outgoing parton of the matrix element that contributes most of the jet constituents.  In this way, at parton level the initiating parton is uniquely defined. At hadron level, however, this is not always the case, as hadrons can't always be associated with only one of the initiating partons. It is thus expected that in some events it will not be possible to make a correspondence between the outgoing parton of the matrix element and the final reconstructed jet. 

To overcome such difficulty, we adopted the following procedure (denoted as v2 in Fig.~\ref{fig:compParton} and \ref{fig:compHadron}): 
\begin{itemize}
    \item Identify the outgoing particles of the matrix element
    \item Tag all final state particles produced by such initiating partons\footnote{At hadron level assigning a hadron to an initiating parton is only an approximation, but we found that it works satisfactorily for our purpose.}
    \item Use them as input to reconstruct anti-\kt jets with $R = 1$.
    \item Compare the $\Delta R$ distance between the leading jet and the two jets obtained in this way.
    \item Identify the initiating parton as the one whose final jet is the closest one in $\Delta R$ (with up to $\Delta R < 1$).
\end{itemize}

In \pythia at hadron level this procedure yields a loss of $9\%$ of the selected events (with at least one jet with $p_{T} > 300$~GeV and $\eta < 1$), compared to the $19\%$ when adopting the approach of going through the ancestor list. 

We compared the obtained $\Delta \tau$ distributions, at parton and hadron level, following these two procedures, and the results are statistically consistent (see Fig.~\ref{fig:compTime}). At parton level, the number of lost events is about $1\%$ for both v1 and v2 $2\%$ at hadron level. Both tagging procedures (v1 and v2) fail to identify $1\%$ of the events at parton level and $2\%$ of the events at hadron level.

In \jwl, however, such procedures are not possible, as the hadronization vertex links all particles of the hard scattering event. Therefore, we instead reconstruct $R=1$ jets from all final state partons (instead of hadrons) originated from the matrix element. This reduces the effects of the hadronization and increase the success rate of identifying the jet initiating parton. In addition, for \jwl (PbPb), we also follow the outgoing medium particles that participate in the elastic scattering process. Overall, these modifications introduce a loss of $2\%$ of the selected events in \jwl (pp) and \jwl (PbPb). As such, we expect the obtained final distributions of $\Delta \tau$ to be unbiased by the tracing procedure.

\section{Correlation as a function of the jet clustering radius and $z_{cut}$}
\label{app:evolSettings}

\begin{figure}[h!]
\centering
\subfloat[Reclustering with C/A algorithm]{
\centering
\includegraphics[width=\columnwidth]{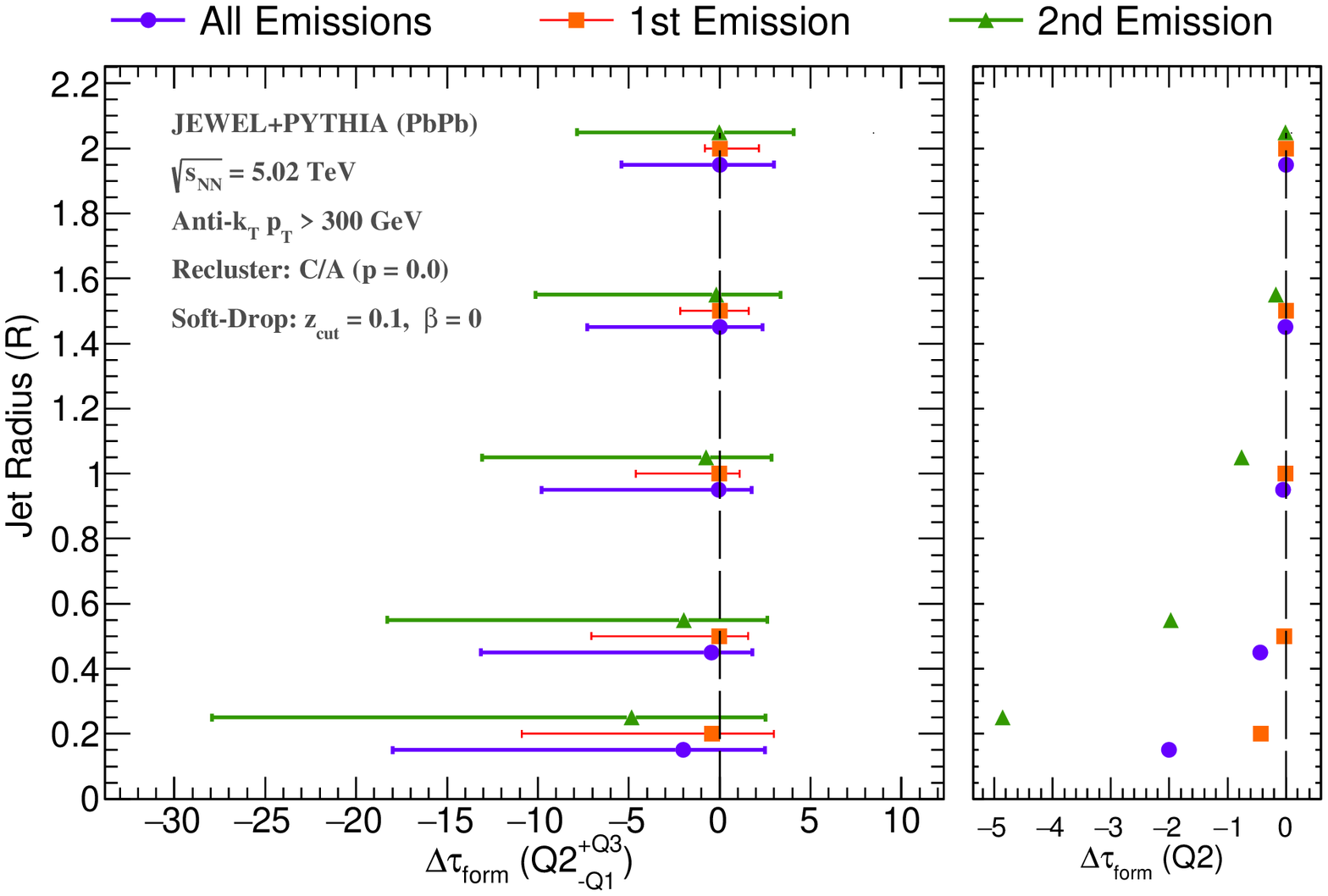}
\label{fig:evolRJWL_ca}}
\linebreak
\subfloat[Reclustering with $\tau$ algorithm]{
\centering
\includegraphics[width=\columnwidth]{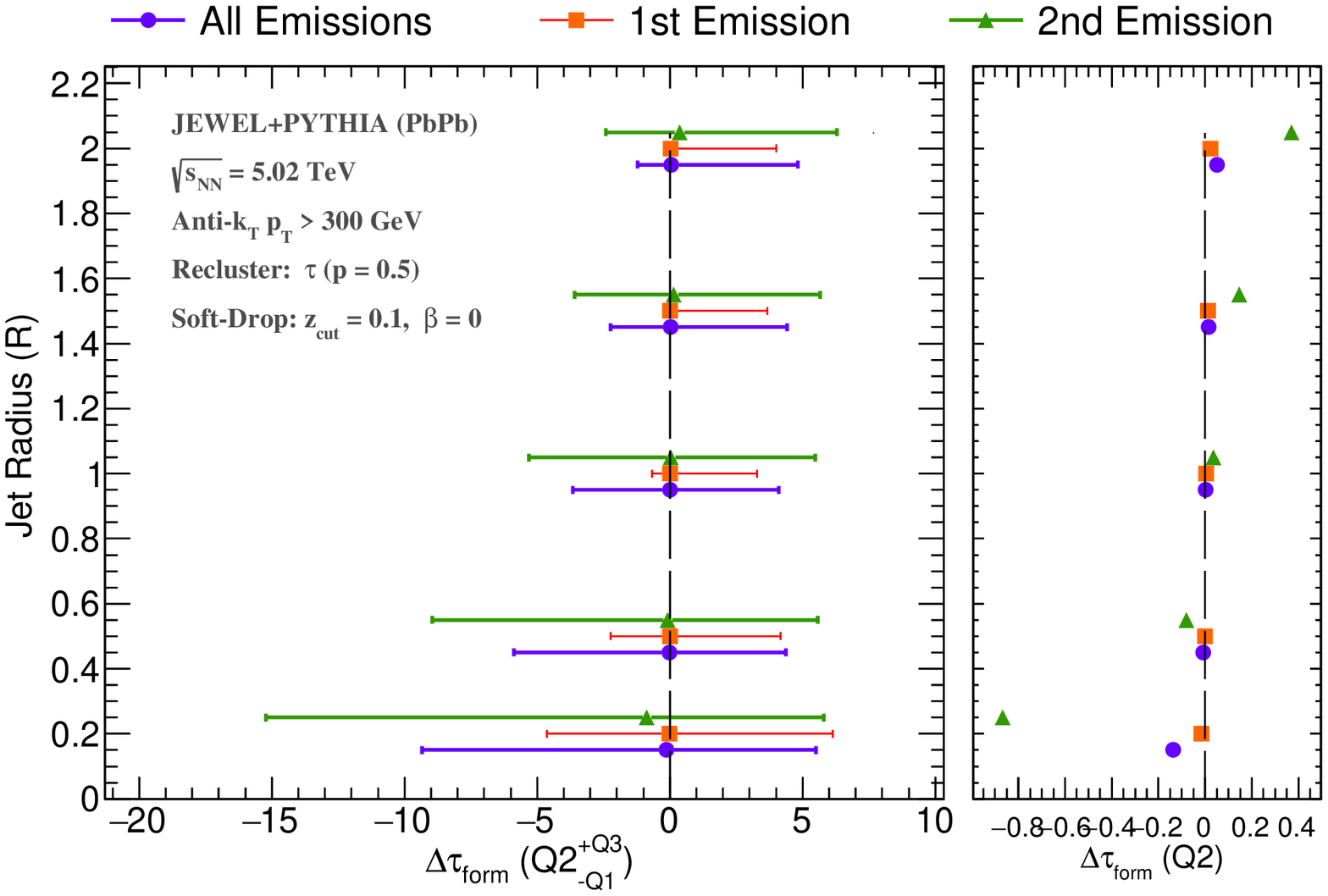}
\label{fig:evolRJWL_tau}}
\caption{Median value of the $\Delta \tau$ distribution (Eq.~\eqref{eq:DeltaTau}) for different jet radii for the first (orange), second (green) or all emissions along the primary branch (purple) in \jwl (PbPb). The asymmetric error bars correspond to the $+Q_3$ and $-Q_1$ quartiles. On the right, a zoom of $Q_2$ alone is shown.}
\label{fig:evolRJWL}
\end{figure}

\begin{figure}[h!]
\centering
\subfloat[Reclustering with C/A algorithm]{
\centering
\includegraphics[width=\columnwidth]{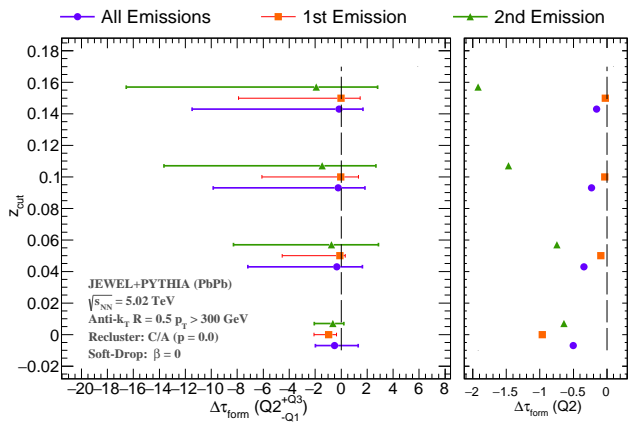}
\label{fig:evolzcutJWL_ca}}
\linebreak
\subfloat[Reclustering with $\tau$ algorithm]{
\centering
\includegraphics[width=\columnwidth]{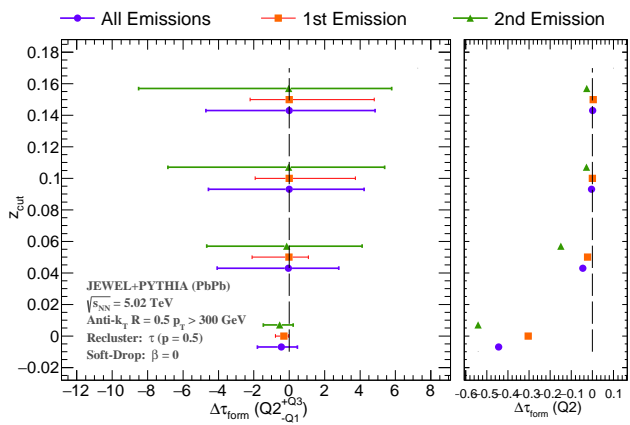}
\label{fig:evolzcutJWL_tau}}
\caption{Median value of the $\Delta \tau$ distribution (Eq.~\eqref{eq:DeltaTau}) obtained with different values of the SoftDrop parameter \zcut for the first (orange), second (green) or all emissions along the primary branch (purple) in \jwl (PbPb). The asymmetric error bars correspond to the $+Q_3$ and $-Q_1$ quartiles. On the right, a zoom of $Q_2$ alone is shown.}
\label{fig:evolzcutJWL}
\end{figure}

The equivalent of Fig.~\ref{fig:evolpJWL}, but fixing $p = 0$ (C/A) or $p = 0.5$ ($\tau$) while varying the jet radius, $R$, and the SoftDrop parameter \zcut, are shown in Fig.~\ref{fig:evolRJWL} and Fig.~\ref{fig:evolzcutJWL}. 

As expected, an increase of the jet radius allows to capture more particles coming from the parton shower, leading to a more complete reconstruction of the jet fragmentation pattern. The matching between parton shower and reclustering algorithms is thus improved, up to $R = 1.0$. From there, the results seem to stabilize, and there is no significant gain from increasing the jet radius further. We would like to note that increasing the jet radius in PbPb events is experimentally challenging given the large multiplicity~\cite{CMS:2019btm,ATLAS:2019rmd}. Moreover, one would also expect an increasing contamination from initial state radiation (ISR). In our case, we do not expect a large effect from ISR given the central rapidity range ($|\eta| < 1$), but this effect should be taken also into account if a larger $\eta$ range is desired. 

The dependence on \zcut is the less obvious. By increasing this parameter, the resulting $\Delta \tau$ distribution for the first emission is shifted towards values near zero, but the overall dispersion is increased. Two competing effects are at play. Introducing a \zcut allows to select only hard QCD emissions, that are more easily matched between clustering and parton shower history (the correlation between the two procedures improves, as observed in section \ref{sec:corr}). On the other hand, even though the transverse momentum selection is made on the ungroomed jet $p_T$, grooming also introduces a selection bias favouring jets with a harder fragmentation, and thus larger formation times. When there is a mismatch between the estimated \taf obtained from reclustering and parton shower history, the magnitude of the possible $\Delta\tau$ is thus also increased. Overall, we find that a \zcut broadens the distribution, but also shifts it to $\Delta \tau\sim0$.

\section{Effect of the \taf definition for jet quenching studies}
\label{app:tau12}

When calculating \taf from Eq.~\eqref{eq:tau} (rightmost expression) an approximation is used. We found that if instead we use
\beq
\tau_{form} \approx \frac{p_{T,1}+p_{T,2}}{p_{T,1} \, p_{T,2} \, \Delta R^2 } \,
\label{eq:tauA}
\eeq
where $p_{T,i}$ refers to the transverse momentum of subjet $1$ and $2$ and $\Delta R^2$ to the $(y,\phi)$ distance between the two, the results are very similar to the ones obtained from Eq.~\eqref{eq:tau}. Both expressions assume high-energy limit, and neglect the virtuality (mass) of the parton (jet). In the parton shower, the assumption of massless particles might be unrealistic, in particular, if we focus on the first emission. Therefore, we have studied the impact on our distributions if, instead, we use
\beq
\tau_{form} \approx \frac{E}{Q^2} = \frac{E}{m_1^2 + m_2^2 + 2 p_1 \cdot p_2} \, ,
\label{eq:tau2}
\eeq
where $m_{1}$ and $m_{2}$ now refers to the virtuality (mass) of the 2 outgoing partons (subjets), and $p_{i}$ to their respective 4-momenta.

\begin{figure}[h!]
\centering
\includegraphics[width=.9\columnwidth]{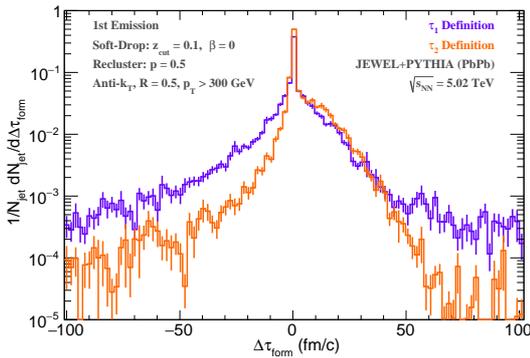}
\caption{$\Delta \tau$ distribution as defined in Eq.~\eqref{eq:DeltaTau} with \taf defined by Eq.~\eqref{eq:tau} (in purple) and Eq.~\eqref{eq:tau2} (in orange), for the first emission/unclustering step with the \ta jet algorithm ($p = 0.5$), and $z_{cut} = 0.1$. }
\label{fig:deltaTau_tau1v2_p05}
\end{figure}

\begin{figure}[h!]
\centering
\includegraphics[width=.9\columnwidth]{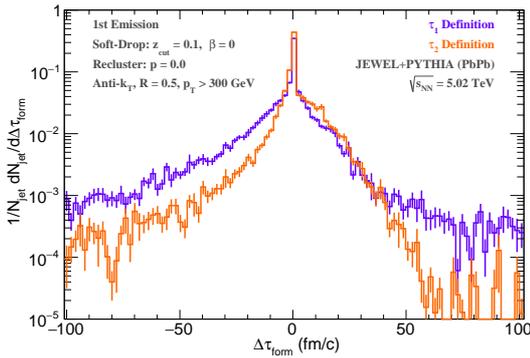}
\caption{$\Delta \tau$ distribution as defined in Eq.~\eqref{eq:DeltaTau} with \taf defined by Eq.~\eqref{eq:tau} (in purple) and Eq.~\eqref{eq:tau2} (in orange), for the first emission/unclustering step with the \ca jet algorithm ($p = 0.0$), and $z_{cut} = 0.1$.}
\label{fig:deltaTau_tau1v2_p00}
\end{figure}

Focusing only on the first emission, we show in Fig.~\ref{fig:deltaTau_tau1v2_p05} the comparison between the $\Delta \tau$ distribution when using \taf as defined by equation \eqref{eq:tau} (named $\tau_1$, in purple) and when we use the definition in \eqref{eq:tau2} (named $\tau_2$, in orange). For both the \ta algorithm is used in the reclustering procedure. By comparison, $\tau_2$ (eq.~\eqref{eq:tau2}) produces a more asymmetric distribution, typically yielding a smaller \ta with respect to the parton shower. This is not entirely unexpected, since $\tau_2$ involves the sub-jet masses, a quantity that is known to receive large non-perturbative corrections. The sub-jet invariant mass obtained at hadron level is therefore not a reliable estimate of the parton level quantity, i.e.\ the parton's virtuality. Since the virtualities are typically small compared to the parton energy, it is safer to assume the sub-jets to be massless. Under this assumption equation~\eqref{eq:tau2} reduces to equation~\eqref{eq:tau}.
The same effect is visible when using the \ca algorithm (Fig.~\ref{fig:deltaTau_tau1v2_p00}), but the shift seems to be less pronounced.
Based on these results, we conclude that $\tau_2$ is, although formally more precise, less robust and decided to keep the definition of \taf as provided by eq.\eqref{eq:tau}. Still, this illustrates the importance of future studies to identify an accurate estimator for the formation time of in-medium parton shower emissions.

\bibliographystyle{spphys}       
\bibliography{Bibliography.bib}   

\begin{thebibliography}{10}
\providecommand{\url}[1]{{#1}}
\providecommand{\urlprefix}{URL }
\expandafter\ifx\csname urlstyle\endcsname\relax
  \providecommand{\doi}[1]{DOI \discretionary{}{}{}#1}\else
  \providecommand{\doi}{DOI \discretionary{}{}{}\begingroup
  \urlstyle{rm}\Url}\fi

\bibitem{Salam:2009jx}
G.P. Salam, Eur. Phys. J. C \textbf{67}, 637 (2010).
\newblock \doi{10.1140/epjc/s10052-010-1314-6}

\bibitem{Dokshitzer:1997in}
Y.L. Dokshitzer, G.~Leder, S.~Moretti, B.~Webber, JHEP \textbf{08}, 001 (1997).
\newblock \doi{10.1088/1126-6708/1997/08/001}

\bibitem{Wobisch:1998wt}
M.~Wobisch, T.~Wengler, in \emph{{Workshop on Monte Carlo Generators for HERA
  Physics (Plenary Starting Meeting)}} (1998), pp. 270--279

\bibitem{Cacciari:2008gp}
M.~Cacciari, G.P. Salam, G.~Soyez, JHEP \textbf{04}, 063 (2008).
\newblock \doi{10.1088/1126-6708/2008/04/063}

\bibitem{Catani:1993hr}
S.~Catani, Y.L. Dokshitzer, M.~Seymour, B.~Webber, Nucl. Phys. B \textbf{406},
  187 (1993).
\newblock \doi{10.1016/0550-3213(93)90166-M}

\bibitem{Ellis:1993tq}
S.D. Ellis, D.E. Soper, Phys. Rev. D \textbf{48}, 3160 (1993).
\newblock \doi{10.1103/PhysRevD.48.3160}

\bibitem{Butterworth:2008iy}
J.M. Butterworth, A.R. Davison, M.~Rubin, G.P. Salam, Phys. Rev. Lett.
  \textbf{100}, 242001 (2008).
\newblock \doi{10.1103/PhysRevLett.100.242001}

\bibitem{Larkoski:2014wba}
A.J. Larkoski, S.~Marzani, G.~Soyez, J.~Thaler, JHEP \textbf{05}, 146 (2014).
\newblock \doi{10.1007/JHEP05(2014)146}

\bibitem{Larkoski:2015lea}
A.J. Larkoski, S.~Marzani, J.~Thaler, Phys. Rev. D \textbf{91}(11), 111501
  (2015).
\newblock \doi{10.1103/PhysRevD.91.111501}

\bibitem{Sirunyan:2017bsd}
A.M. Sirunyan, et~al., Phys. Rev. Lett. \textbf{120}(14), 142302 (2018).
\newblock \doi{10.1103/PhysRevLett.120.142302}

\bibitem{Caucal:2019uvr}
P.~Caucal, E.~Iancu, G.~Soyez, JHEP \textbf{10}, 273 (2019).
\newblock \doi{10.1007/JHEP10(2019)273}

\bibitem{Milhano:2017nzm}
G.~Milhano, U.A. Wiedemann, K.C. Zapp, Phys. Lett. B \textbf{779}, 409 (2018).
\newblock \doi{10.1016/j.physletb.2018.01.029}

\bibitem{Andrews:2018jcm}
H.A. Andrews, et~al.,   (2018)

\bibitem{Dokshitzer:1991wu}
Y.L. Dokshitzer, V.A. Khoze, A.H. Mueller, S.~Troian, \emph{{Basics of
  perturbative QCD}} (1991)

\bibitem{CasalderreySolana:2012ef}
J.~Casalderrey-Solana, Y.~Mehtar-Tani, C.A. Salgado, K.~Tywoniuk, Phys. Lett. B
  \textbf{725}, 357 (2013).
\newblock \doi{10.1016/j.physletb.2013.07.046}

\bibitem{CasalderreySolana:2011rz}
J.~Casalderrey-Solana, E.~Iancu, JHEP \textbf{08}, 015 (2011).
\newblock \doi{10.1007/JHEP08(2011)015}

\bibitem{Sjostrand:2006za}
T.~Sjostrand, S.~Mrenna, P.Z. Skands, JHEP \textbf{05}, 026 (2006).
\newblock \doi{10.1088/1126-6708/2006/05/026}

\bibitem{Bothmann:2019yzt}
E.~Bothmann, et~al., SciPost Phys. \textbf{7}(3), 034 (2019).
\newblock \doi{10.21468/SciPostPhys.7.3.034}

\bibitem{Bellm:2015jjp}
J.~Bellm, et~al., Eur. Phys. J. C \textbf{76}(4), 196 (2016).
\newblock \doi{10.1140/epjc/s10052-016-4018-8}

\bibitem{Sjostrand:2007gs}
T.~Sjostrand, S.~Mrenna, P.Z. Skands, Comput. Phys. Commun. \textbf{178}, 852
  (2008).
\newblock \doi{10.1016/j.cpc.2008.01.036}

\bibitem{Zapp:2013vla}
K.C. Zapp, Eur. Phys. J. C \textbf{74}(2), 2762 (2014).
\newblock \doi{10.1140/epjc/s10052-014-2762-1}

\bibitem{Corke:2010yf}
R.~Corke, T.~Sjostrand, JHEP \textbf{03}, 032 (2011).
\newblock \doi{10.1007/JHEP03(2011)032}

\bibitem{Bjorken:1982qr}
J.~Bjorken, Phys. Rev. D \textbf{27}, 140 (1983).
\newblock \doi{10.1103/PhysRevD.27.140}

\bibitem{Acharya:2018qsh}
S.~Acharya, et~al., JHEP \textbf{11}, 013 (2018).
\newblock \doi{10.1007/JHEP11(2018)013}

\bibitem{Aaboud:2018twu}
M.~Aaboud, et~al., Phys. Lett. B \textbf{790}, 108 (2019).
\newblock \doi{10.1016/j.physletb.2018.10.076}

\bibitem{CMS:2019btm}
{CMS Collaboration},   (2019)

\bibitem{paperKorinnaNew}
J.G. Milhano, K.~Zapp, Under Preparation

\bibitem{Cacciari:2011ma}
M.~Cacciari, G.P. Salam, G.~Soyez, Eur. Phys. J. C \textbf{72}, 1896 (2012).
\newblock \doi{10.1140/epjc/s10052-012-1896-2}

\bibitem{ATLAS:2019rmd}
{The ATLAS collaboration},   (2019)

\bibitem{Apolinario:2020zvt}
L.~Apolin\'{a}rio, in \emph{{2019 European Physical Society Conference on High
  Energy Physics}} (2020)

\bibitem{Citron:2018lsq}
Z.~Citron, et~al., \emph{{Report from Working Group 5}: {Future physics
  opportunities for high-density QCD at the LHC with heavy-ion and proton
  beams}} (2019), vol.~7, pp. 1159--1410.
\newblock \doi{10.23731/CYRM-2019-007.1159}

\bibitem{Apolinario:2017sob}
L.~Apolin\'{a}rio, J.G. Milhano, G.P. Salam, C.A. Salgado, Phys. Rev. Lett.
  \textbf{120}(23), 232301 (2018).
\newblock \doi{10.1103/PhysRevLett.120.232301}

\bibitem{Andres:2019eus}
C.~Andres, N.~Armesto, H.~Niemi, R.~Paatelainen, C.A. Salgado, Phys. Lett. B
  \textbf{803}, 135318 (2020).
\newblock \doi{10.1016/j.physletb.2020.135318}

\bibitem{CasalderreySolana:2010eh}
J.~Casalderrey-Solana, J.G. Milhano, U.A. Wiedemann, J. Phys. G \textbf{38},
  035006 (2011).
\newblock \doi{10.1088/0954-3899/38/3/035006}

\bibitem{Apolinario:2017qay}
L.~Apolin\'ario, J.G. Milhano, M.~Ploskon, X.~Zhang, Eur. Phys. J. C
  \textbf{78}(6), 529 (2018).
\newblock \doi{10.1140/epjc/s10052-018-5999-2}

\end{thebibliography}

\end{document}